\documentclass{aastex}

\usepackage{emulateapj5}
\newcommand{\lsim}{\mbox{\hspace{.2em}\raisebox{.5ex}{$<$}\hspace{-.8em}\raisebox{-.5ex}{$\sim$}\hspace{.2em}}}
\newcommand{\gsim}{\mbox{\hspace{.2em}\raisebox{.5ex}{$>$}\hspace{-.8em}\raisebox{-.5ex}{$\sim$}\hspace{.2em}}}

\def\asca       {{\em ASCA}\/}
\def\chandra    {{\em Chandra}\/}

\def\xmm        {{\em XMM-Newton}\/}
\def\xmma       {{\em XMM}\/}
\def\rosat      {{\em ROSAT}\/}

\def\mydegree{$^\circ\mskip-5mu$}

\def\ga         {{ESO~137-001}\/}

\begin{document}

\title{A 70 kpc X-ray tail in the cluster A3627}

\author{
M.\ Sun,$^{\!}$\altaffilmark{1}
C.\ Jones,$^{\!}$\altaffilmark{2}
W.\ Forman,$^{\!}$\altaffilmark{2}
P.\ E.\ J.\ Nulsen,$^{\!}$\altaffilmark{2,3}
M.\ Donahue,$^{\!}$\altaffilmark{1} 
M.\ Voit$^{\!}$\altaffilmark{1}
}
\smallskip

\affil{\scriptsize 1) Department of Physics and Astronomy, MSU, East Lansing, MI 48824; sunm@pa.msu.edu}
\affil{\scriptsize 2) Harvard-Smithsonian Center for Astrophysics,
60 Garden St., Cambridge, MA 02138}
\affil{\scriptsize 3) University of Wollongong, NSW 2522, Australia, on leave} 	

\shorttitle{An X-ray tail in A3627}
\shortauthors{Sun et al.}

\begin{abstract}

We present the discovery of a 70 kpc X-ray tail behind the
small late-type galaxy \ga\ in the nearby, hot (T=6.5 keV)
merging cluster A3627, from both \chandra\ and \xmm\
observations. The tail has a length-to-width ratio of $\sim$
10. It is luminous (L$_{\rm 0.5-2 keV} \sim$ 10$^{41}$ ergs
s$^{-1}$), with a temperature of $\sim$ 0.7 keV and an X-ray
M$_{\rm gas}$ of $\sim 10^{9}$ M$_{\odot}$ ($\sim$ 10\% of the
galaxy's stellar mass). We interpret this tail as the stripped
interstellar medium of \ga\ mixed with the hot cluster medium,
when this blue galaxy is being converted into a gas-poor galaxy.
Three X-ray point sources are detected in the axis of the tail,
which may imply active star formation in the tail. The straightness
and narrowness of the tail also implies that the ICM turbulence
is not strong on scales of 20 - 70 kpc. 

\end{abstract}

\keywords{galaxies: clusters: general --- galaxies: clusters: individual
  (A3627) --- X-rays: galaxies --- galaxies: individual (ESO 137-001)}

\section{Introduction}

The intracluster medium (ICM) has long been proposed to play a vital role in
shaping the properties of cluster galaxies, especially through stripping
cold galactic material. Ram-pressure stripping is likely an important
factor for galaxy transformation in rich environments (e.g., Gunn \&
Gott 1972; Quilis, Moore \& Bower 2000). Radio observations have long revealed
an HI deficiency in cluster galaxies, which indicates stripping (e.g., Giovanelli
\& Haynes 1985). In X-rays, the stripped tails of late-type galaxies only began
to be revealed by \chandra\ and \xmm\ data (Wang et al. 2004; Sun \& Vikhlinin
2005, SV05 hereafter; Machacek et al. 2005). Recently we carried out a systematic
analysis of the X-ray thermal coronae of $\sim$ 200 galaxies
in 25 nearby (z $<$ 0.05) hot (kT $>$ 3 keV) clusters (Sun et al. in prep.).
During our analysis, we found an X-ray tail in A3627, which
is apparently associated with the late-type galaxy \ga.
The narrowness and length of the tail makes it the most dramatic
X-ray stripped tail of a galaxy to date.

A3627 is a nearby massive galaxy cluster in the core of the
Great Attractor. Kraan-Korteweg et al. (1996) found, in a deep
imaging and spectroscopic survey for galaxies behind the Milky Way, that
A3627 (at a Galactic latitude of -7.2\mydegree) is an extremely rich cluster
rivaling Coma and Perseus in mass and galactic content. A major cluster merger
is implied by the earlier X-ray observations from \rosat\ and \asca\ that
reveal a southeast - northwest cluster elongation
(Fig.~1a) and a temperature gradient (5 - 8 keV) along the same direction
(B\"ohringer et al. 1996; Tamura et al. 1998).

The radial velocity of \ga\ is 4630 $\pm$ 58 km~s$^{-1}$, almost the same as A3627's,
4707 km~s$^{-1}$ (NASA/IPAC Extragalactic Database - NED). In this paper,
we adopt the cluster redshift to calculate the distance of \ga.
Assuming H$_{0}$=70 km s$^{-1}$ Mpc$^{-1}$, $\Omega$$_{\rm M}$=0.3,
and $\Omega_{\rm \Lambda}$=0.7, the luminosity distance is
68.2 Mpc, and 1$''$=0.321 kpc. Uncertainties quoted in this paper are 1 $\sigma$.

\section{\chandra\ and \xmm\ data}

The tailed X-ray source lies in a 15 ks \chandra\ observation of A3627 (ObsID 4956),
performed with the Advanced CCD Imaging Spectrometer (ACIS) on June 14 - 15,
2004. No background flares were detected. Standard data analysis has
been performed. The calibration files used correspond to \chandra\
Calibration Database 3.1.0.

The source is also contained within an 18 ks \xmm\ observation on August 12,
2004. The data have been reduced with the Science Analysis Software 6.5.0.
After removing large background flares, clean exposures of 17.8 ks,
18.0 ks and 12.1 ks remain for 3 CCD cameras MOS1, MOS2 and PN respectively.

\section{The tailed X-ray source}

The \xmm\ photon image and adaptively smoothed \chandra\ images are shown
in Fig.~1b, c and d, which clearly show a long narrow tail. The tail
is only visible in the soft X-ray images. Its brighter southeast end
is on the cluster galaxy \ga\ (Fig. 1c). There are no
other galaxies as bright as \ga\ in the B band within 7$'$ of \ga\ in the
direction of the tail. Thus, we conclude the X-ray extended source is physically
associated with \ga.

\ga\ is in a crowded field with many Galactic stars.
The galaxy is bright in the optical B band,
but faint in the 2MASS Ks band. The B band luminosity is 2.3-4.0$\times10^{10}$
L$_{\rm B\odot}$ from NED and HyperLeda, depending on the correction for internal
extinction. The Ks band luminosity is 1.4$\times10^{10}$ L$_{\rm K\odot}$. Assuming a Ks
band mass-to-light ratio of 0.73 M$_{\odot}$/L$_{\odot}$ (Cole et al. 2001),
the total stellar mass of \ga\ is only 1.1$\times10^{10}$ M$_{\odot}$, which
makes it $\sim$ 5 times less massive than a L$_{*}$ galaxy.
The B-Ks color of \ga\ is 1.0-1.6, much smaller than the B-Ks color of a typical
early-type or late-type galaxy (3.7-4.0 or $\sim$ 3, Jarrett 2000).
Thus, \ga\ is a very blue galaxy. Spectroscopic studies (Woudt, Kraan-Korteweg \&
Fairall 1999; Woudt et al. 2004) revealed emission lines of H$\alpha$, H$\beta$,
[OIII] 5007 and [NII] 6584 in the spectrum of \ga. Thus, we conclude that \ga\
is a late-type galaxy with active star formation.

\subsection{The X-ray diffuse emission}

To better quantify the
properties of the tail, we measured \chandra\ surface brightness
profiles along and across it (Fig. 1e and f). The surface brightness
has been corrected for background (from the blank sky background) and
exposure. There is a sharp edge at its southeast end with a
brightness jump of a factor of $\sim$ 3. The tail is detected to 3.7$'$
(or 71 kpc) northwest of \ga. The width of the tail is $\lsim 25''$ (or 8 kpc).
If we fit the profile across the tail with a Gaussian plus the local background,
the derived Full Width Half Maximum (FWHM) is only 15.2$^{+2.3}_{-2.1}$$''$.
Thus, this feature is very narrow with a length-to-width ratio of $\sim$ 10.

Integrated \chandra\ and \xmm\ spectra of the source were extracted from a
box region centered on the source (3.7$'\times30''$ for \chandra\ and 4.1$'\times50''$
for \xmma). We were forced to use a larger extraction region for \xmma\ as the
width of the tail is comparable to the Half Power Diameter of the \xmma\ mirror
(15$''$). Background was taken from surrounding regions, a 4.5$'\times2'$
box for \chandra\ and 5$'\times2.5'$ box for \xmma, excluding the source region.
Within the source region, 230$\pm$24 (point sources excluded), 967$\pm$95
and 872$\pm$103 net counts in the 0.5 - 2 keV band are collected from
the ACIS, PN and MOS instruments respectively.
The current data do not allow us to constrain the absorption. As the
spectra are well fitted with Galactic absorption at low energies,
we fixed the absorption at the Galactic value, 2$\times$10$^{21}$ cm$^{-2}$.
The solar photospheric abundance table by Anders \& Grevesse (1989) is adopted.
A lower energy cut of 0.5 keV is used to minimize the calibration uncertainties
at low energies. The upper energy cut is 7 keV. Each spectrum is fitted to the
MEKAL model. The abundance is very low but the gas
temperature is not sensitive to abundance (e.g., T=0.66$\pm$0.06 if Z=1 solar).
All spectra showed evidence for a hard X-ray excess so a power law component
was included. Inclusion of the hard component lowers the gas temperature
by 10\% - 20\%, but greatly improves the fit. The nature of this hard X-ray
component is unclear. The expected unresolved AGN contribution is small.
There may be some residual emission from the hot ICM, either mixed in the
tail or due to the uncertain normalization of the background. Emission of
unresolved X-ray binaries in the tail (see $\S$3.2) is another possibility.
Regardless of the model used for the hard component, hot ICM (MEKAL), X-ray
binaries, or AGN (power law), the results for gas properties are almost the same.
The \chandra\ and \xmma\ results are consistent so we fit them simultaneously.
The gas temperature is $\sim$ 0.7 keV (Table 1). We also derived temperatures
in two regions, Head and Trail, defined in Fig. 1f. The Trail
is only marginally hotter (1 $\sigma$) than the Head.
There are some discrepancies between the derived \chandra\ and \xmma\ luminosities,
at the level of 20\% - 33\%, which may be due to uncertainties on the background
normalization (especially for \xmma, where the background is higher).

The X-ray gas mass in the tail is estimated by assuming a 25$''\times25''\times1.1'$
cylinder for the Head and a 2.6$'$ long cylinder for the Trail with constant
density. From the normalization derived from the MEKAL fits,
the average electron densities in the Head and the Trail are
1.8-3.5 $\times 10^{-2}$ f$^{-1/2}$ cm$^{-3}$ and 0.7-1.7
$\times 10^{-2}$ f$^{-1/2}$ cm$^{-3}$ respectively, while the X-ray gas mass is
0.7-1.3 $\times 10^{9}$ f$^{1/2}$ M$_{\odot}$ and 0.6-1.5 $\times 10^{9}$
f$^{1/2}$ M$_{\odot}$ respectively, where f is the filling factor of the
X-ray gas. If the abundance in Head is 0.3 solar (the value
in the Trail within 1 $\sigma$), the average electron density and the X-ray
gas mass of the Head drop by 40\%. The filling factor may be large if the
observed tail is really due to mixing of cold clouds and hot ICM (see $\S$4).
Since the galaxy and cluster radial velocity are nearly the same, the projected
length of the tail may be comparable to the actual length.
Thus, the total X-ray gas mass is on the order of 10$^{9}$ M$_{\odot}$.

The temperature of the surrounding ICM (a 3$'\times6'$ box, excluding the tailed
source) is also measured. Since A3627 is behind the Galactic plane, the
Galactic soft X-ray foreground is strong. This is proved from the 3 - 5 times
higher than nominal flux in the R4-R5 band of the RASS image at the position of
A3627. As there is no local
background available, we used the blank-sky background and added a 0.2 keV
MEKAL component (unabsorbed) to mimic the soft X-ray excess. The derived ICM
temperature is 6.3$^{+1.1}_{-0.7}$ keV (Table 1). The error from the uncertainty
on the temperature of the soft excess is included, as we allow it to vary from
0.1 to 0.4 keV (Markevitch et al. 2003).
The ICM electron density from the \rosat\ data (B\"ohringer et al. 1996) is $\sim$
1.4$\times10^{-3}$ cm$^{-3}$ at the projected position of \ga. Within uncertainties,
the thermal pressure is the same inside and outside the tail, especially if the
abundance is not that low.

\subsection{X-ray point sources in the tail}

There are three \chandra\ point sources right on the axis of the tail
(P1 - P3 in Fig. 1), with a total of 73 counts in the 0.5 - 10 keV band.
Their combined spectrum is hard (Table. 1).
Within 3$'$ of \ga, only an additional, fainter point source is detected.
The possibility of having three background AGN aligned in the narrow tail
for this shallow observation is $\sim$ 0.03\%, based on the AGN number density
derived from three \chandra\ pointings of A3627. If P1-P3 are located in
the tail, L$_{\rm 2 - 10 keV}$ = 2.7 $\times 10^{40}$ ergs s$^{-1}$ (the total),
which is similar to the total L$_{\rm X}$ of three ULXs in UGC~6697 (SV05). Since
the ICM pressure cannot move stars, the most likely scenario is that there is
active massive star formation in the tail, as the stripped clouds, away from
their main heating source (the stellar UV radiation), can cool. H$\alpha$
imaging and HI data are required to test this hypothesis.

\section{Discussion}

The X-ray tail of \ga\ is unique. Our systematic study covers
a sky area of 3.1 deg$^{2}$ in 25 nearby hot clusters, but only UGC~6697
in A1367 has a similar tail as \ga. However, UGC~6697 is 8 times more
luminous than \ga\ in the Ks band and its tail does not extend significantly
outside of the galaxy (SV05). Even the stripped X-ray tails known in cool
clusters and groups (e.g., Machacek et al. 2005) are not as dramatic as
the tail of \ga. The X-ray tail of \ga\ is only the third known X-ray
tail of a late-type galaxy in hot clusters (after Wang et al. 2004 and SV05).
While its proximity enables further detailed studies, the tail length and the
X-ray point sources in the tail provide additional insights.

Long and narrow X-ray tails have been seen in the simulation by
Stevens, Acreman \& Ponman (1999), when the stripping happens gradually and
gas is replenished in the galaxy. Similar HI tails
also were shown in other simulations (e.g., Quilis et al. 2000; Vollmer et al. 2001).
Two 75 kpc H$\alpha$ trails have been detected behind two small late-type
galaxies in A1367 (Gavazzi et al. 2001), but no X-ray tails are seen in the
\xmma\ data. A 110 kpc but tilted HI tail was observed from NGC~4388
in Virgo (Oosterloo \& van Gorkom 2005).	
The tail of \ga\ cannot be the accretion wake of \ga, as the galaxy is too
small and the ICM is too hot. The \ga\ tail can either be
the stripped X-ray gas of \ga, or the cold
material of \ga\ (e.g., HI gas) mixed with hot ICM.
The first scenario requires that the stripped tenuous X-ray gas survives
all kinds of instabilities and yet remains distinct from the surrounding ICM
for $\sim 10^{8}$ yr, which is about the time needed for \ga\ to travel from
the end of the tail to its current position (71 kpc in projection) with a velocity
of 1000 km~s$^{-1}$ (the cluster radial velocity dispersion is 897 km~s$^{-1}$).
Mixing of cold ISM with hot ICM is preferred, as it is expected after
stripping of cold ISM. The average temperature after mixing ($\sim$ 0.7 keV) is:
(M$_{\rm ICM}$T$_{\rm ICM}$+M$_{\rm ISM}$T$_{\rm ISM}$)/(M$_{\rm ICM}$+M$_{\rm ISM}$),
which gives M$_{\rm ISM}$ /
M$_{\rm ICM} \sim$ 8 for T$_{\rm ICM}$=6.3 keV. Thus, the cold ISM dominates
the mass but resides in a much smaller volume than the hot ICM before the mixing,
if pressure equilibrium is assumed. The evaporation rate of the X-ray tail by the
hot ICM is unconstrained within the uncertainties, especially if the mixing
is still replenishing the $\sim$ 0.7 keV X-ray tail.

As \ga\ is not a massive galaxy, it should not be hard to remove its gas. 
According to the approximate criterion for ram pressure stripping by Gunn \&
Gott (1972), at a velocity of $\sim$ 800 km~s$^{-1}$, ISM with a density of
1 cm$^{-3}$ can be removed, if n$_{\rm ICM}=10^{-3}$ cm$^{-3}$. From the 
approximation by Mori \& Burkert (2000), the stripping process may take
$\gsim 10^{8}$ yr, if the ISM density is $>$ 1 cm$^{-3}$. Thus, the timescale
is long enough to explain the length of the observed tail, and short enough
to explain the rarity of such long X-ray tails in late-type galaxies.
As there may still be un-mixed cold gas in the tail and galaxy, \ga\ must have
been a very gas-rich galaxy, with a gas fraction of $\gsim$ 0.1. The
star formation rate in \ga\ should be high from its very blue color and emission-line
spectrum. Therefore, \ga\ resembles blue dwarf galaxies in z$\sim$0.4 clusters
that are responsible for the Butcher \& Oemler effect (Butcher \& Oemler 1978).
As there are no other bright cluster galaxies within 7$'$ in the tail direction,
the star formation in \ga\ is likely triggered by the ICM pressure
(e.g., Vollmer et al. 2001; SV05). Thus, \ga, in its first passage through the
cluster core, is being converted into a gas-poor galaxy (likely an E+A galaxy)
after a possible initial starburst, all by the interaction with the dense ICM.

Tails like that of \ga\ may provide constraints on the turbulent velocity field
in clusters.
The ICM has long been suspected to be turbulent, although evidence is sparse.
Turbulence eddies on scales of 20 - 90 kpc have been reported in the Coma
cluster by Schuecker et al. (2004), who concluded that $>$ 10\% of total ICM
pressure is in turbulent form. Dolag et al. (2005) used a novel approach to
treat the ICM viscosity and found that turbulent energy is up to 8\% of the
thermal energy (on a scale of 70 kpc). If we assume that 15\% of the ICM pressure
is turbulent around \ga, the ICM velocity dispersion is $\sim$ 435 km~s$^{-1}$.
Each part of the tail is subject to turbulent pressure in random directions, which
is 2r$\rho_{\rm ICM}$v$_{\rm tur}^{2}$ per unit length, where r is the radius of
the tail cylinder ($\sim$ 4 kpc) and v$_{\rm tur}$ is the turbulent velocity. This
pressure induces an acceleration a=2$\rho_{\rm ICM}$v$_{\rm tur}^{2}$/$\pi$r$\rho_{\rm tail}$
at random directions for different parts of the tail. An eddy on scale of R
lasts for a time of $\sim$ R/v$_{\rm tur}$, which is $\gsim$ 10$^{8}$ yr for R $>$
20 kpc. This implies that large eddies last longer than the current age of the
tail. The displacement of a part of the tail, over a time interval of $t$ is then: 
$at^{2}/2$. The tail is straight and
%$\rho_{\rm ICM}$v$_{\rm tur}^{2}$t$^{2}$ / $\rho_{\rm tail}\pi$r. The tail is straight and
narrow from the end to the head so we set an upper limit of 10$''$ on the
displacement. Assuming thermal pressure equilibrium between the ICM
and the X-ray tail (or $\rho_{\rm tail}$/$\rho_{\rm ICM}$=9 from the temperature ratio),
we have v$_{\rm tur} <$ 380 km~s$^{-1}$ (t/0.05 Gyr)$^{-1}$ on scales from 20 kpc to the
length of the tail (or 70 kpc). This upper limit is still consistent with the lower limit from
Schuecker et al., especially if \ga\ moves very fast.
Turbulence can also be generated as the gas-rich \ga\ moves through the ICM.
However, the Reynold number is $\sim 3{\cal M}(L/\lambda_{\rm i})$, where
${\cal M}$ is the Mach number, $L$ is the size of the galaxy and $\lambda_{\rm i}$
is the mean free path of ions in the ICM. For unmagnetized gas,
$\lambda_{\rm i} \sim$ 10 kpc $\sim$ L. Thus, unless magnetic field has reduced
$\lambda_{\rm i}$ by $>$ 10 times, the motion of \ga\ cannot drive turbulence in the trail.

\acknowledgments

Support for this work was provided by NASA grant NNG-O5GD5Z,
NASA LTSA grant NNG-05GD82G and by the Chandra X-ray center 
and the Smithsonian Institute.

\begin{table}
\begin{center}
\caption{}
{\tiny
\begin{tabular}{lccccc} \hline \hline
Region$^{\rm a}$ & Instrument & Model$^{\rm b}$ & Parameters$^{\rm c}$ & L$_{\rm 0.5 - 2 keV}^{\rm d}$ (gas) & L$_{\rm 0.5 - 2 keV}^{\rm d}$ (hard)\\ \hline

Total & \chandra\ / XMM & MEKAL+PL & T=0.69$\pm$0.08 keV, Z=0.05$^{+0.05}_{-0.02}$ solar & 8.6 (ACIS), 12 (PN), 10 (MOS) & 1.2 (ACIS), 5.7 (XMM) \\
\hspace{0.1cm} Head  & \chandra\ / XMM & MEKAL+PL & T=0.64$^{+0.08}_{-0.22}$ keV, Z=0.04$^{+0.05}_{-0.03}$ solar & 4.0 (ACIS), 5.5 (PN), 4.7 (MOS) & 1.0 (ACIS), 2.1 (XMM) \\
\hspace{0.1cm} Trail  & \chandra\ / XMM & MEKAL+PL & T=0.83$^{+0.13}_{-0.11}$ keV, Z=0.16$^{+0.16}_{-0.10}$ solar & 4.4 (ACIS), 5.7 (PN), 5.4 (MOS) & 0.7 (ACIS), 3.6 (XMM) \\
\hspace{0.1cm} PSs   & \chandra\ & PL       &   $\Gamma$=1.6$\pm$0.4 &  - & 1.3 \\
ICM   & \chandra\ & MEKAL    & T=6.3$^{+1.1}_{-0.7}$ keV, Z=0.1$^{+0.2}_{-0.1}$ solar & - & - \\ 

\hline \hline
\end{tabular}
\vspace{-0.1cm}
\begin{flushleft}
\leftskip 5pt
$^{\rm a}$ See Fig. 1 for the definition of the Head and Trail regions. For
\chandra\ analysis of diffuse emission, three point sources are excluded, while they are
unresolved in the \xmma\ data. The \chandra\ spectra of these three point sources and the
surrounding ICM region ($\S$3.2) are also analyzed.\\
$^{\rm b}$ There is hard excess in each spectrum of the tail, so a power law model (with
a fixed photon index of 1.7) is included in the fits for the Total, Head, and Trail regions.\\ 
$^{\rm c}$ Galactic absorption (2$\times10^{21}$ cm$^{-2}$) is adopted for all
spectral analysis. Cash statistic is used for all spectra except that of the ICM.
Fits are all very good.\\
$^{\rm d}$ The 0.5 - 2 keV luminosities (in units of 10$^{40}$ ergs s$^{-1}$) of the gas and
the hard component \\
\end{flushleft}}
\end{center}
\end{table}

\clearpage

\begin{figure}
\vspace{-2.5cm}
\hspace{-1.0cm}
\includegraphics[scale=0.92]{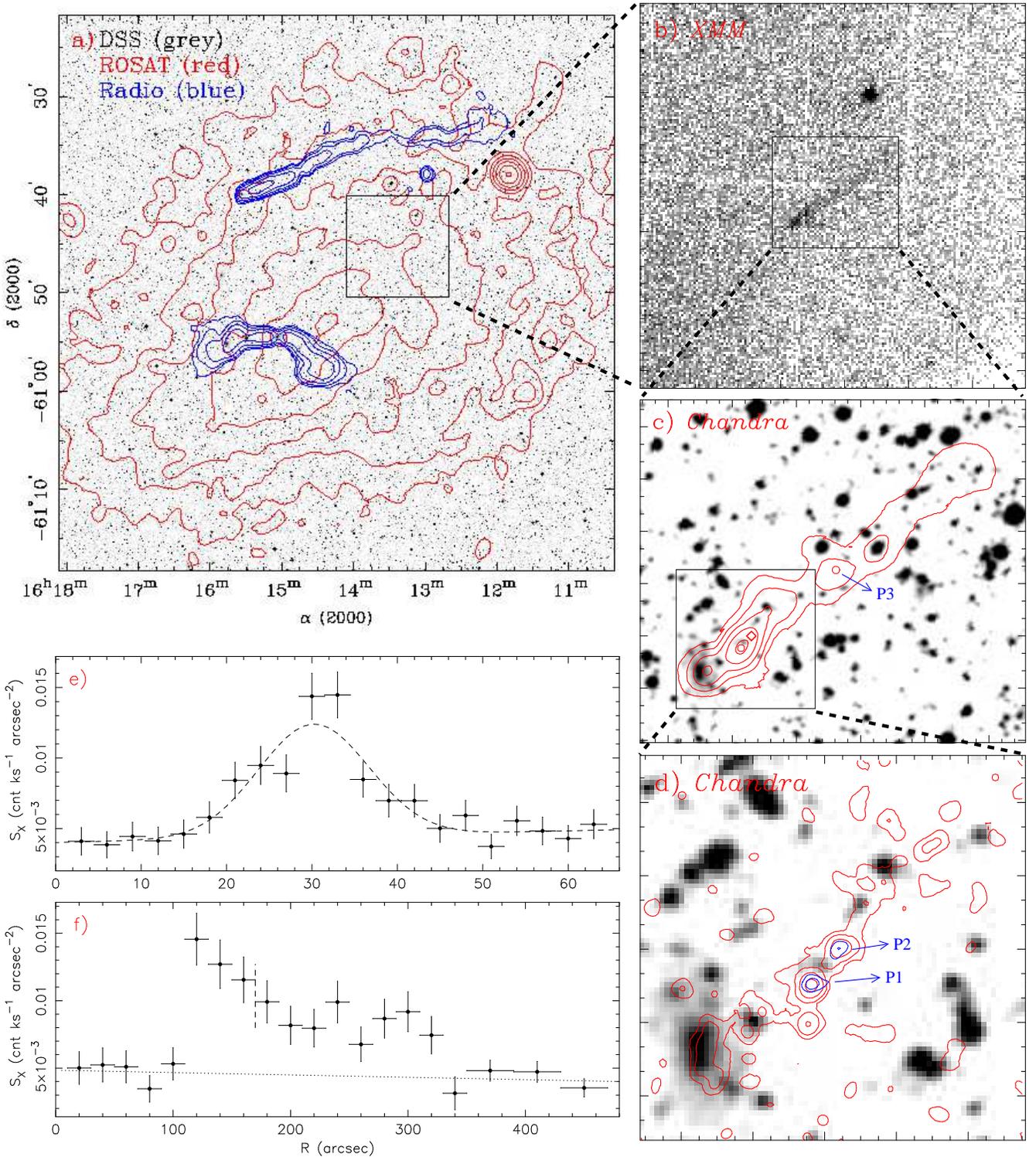}
\vspace{-0.8in}
\caption{
a): \rosat\ contours (red) for A3627 superposed on the DSS image (gray).
The 843 MHz contours of two strong radio sources in A3627 (blue)
also are shown (from the Sydney University Molonglo Sky Survey).
The cluster is in the Galactic plane so the foreground
star density is high. b): \xmma\ 0.5 - 2 keV photon image (PN and
MOS combined, 11.1$'\times11.1'$) of the zoom-in region in a). A
source with a narrow-angle tail, pointing to a similar direction
as the radio tailed source, is significant. c): the zoom-in of b)
(3.7$'\times3.3'$). \chandra\ 0.5 - 2 keV contours are superposed
on DSS II image. The \chandra\ image (background-subtracted and
exposure-corrected) has been adaptively smoothed. The head of the
X-ray tail is on the galaxy \ga. P3 is a point source.
d): the zoom-in of c) (1.3$' \times 1.3'$). The X-ray tail is shown by
the red \chandra\ 0.5 - 2 keV contours on the DSS II image. The 2 - 6 keV contours
of two point sources (P1 and P2) are shown in blue. e): The \chandra\ 0.5 - 2 keV
surface brightness profile across the
tail (from southwest to northeast with a length of 3.7$'$). The dashed line is a
fit to the profile with a Gaussian and background.
f): The \chandra\ 0.5 - 2 keV surface
brightness profile along the direction of the tail (from southeast to northwest
with a width of 25$''$). The dotted line is the fit to the local background, while
the dashed line marks the position where we separate Head and Trail regions in the
spectral analysis. The nucleus of the galaxy is at R=115$''$.
}
\end{figure}


\begin{references}

 \reference{} Anders, E., \& Grevesse N. 1989, Geochimica et Cosmochimica Acta, 53, 197
 \reference{} B\"ohringer, H., Neumann, D. M., Schindler, S., Kraan-Korteweg, R. C. 1996, ApJ, 467, 168
 \reference{} Butcher, H., \& Oemler, A. Jr. 1984, ApJ, 285, 426
 \reference{} Cole, S. et al. 2001, MNRAS, 326, 255
 \reference{} Dolag, K., Vazza, F., Brunetti, G., Tormen, G. 2005, MNRAS, submitted, astro-ph/0507480
 \reference{} Gavazzi, G. et al. 2001, A\&A, 563, L23
 \reference{} Giovanelli, R., Haynes, M. P. 1985, ApJ, 292, 404
 \reference{} Gunn, J. E., \& Gott, J. R., III 1972, ApJ, 176, 1
 \reference{} Jarrett, T. H. 2000, PASP, 112, 1008
 \reference{} Kraan-Korteweg, R. C. et al. 1996, Nature, 379, 519
 \reference{} Machacek, M. E. et al. 2005, ApJ, 630, 280 	
 \reference{} Markevitch, M. et al. 2003, ApJ, 583, 70
 \reference{} Mori, M., \& Burkert, A. 2000, ApJ, 538, 559
 \reference{} Oosterloo, T., van Gorkom, J. 2005, A\&A, 437, 19L
 \reference{} Quilis, V., Moore, B., \& Bower, R. G. 2000, Science, 288, 1617
 \reference{} Schuecker, P. et al. 2004, A\&A, 426, 387
 \reference{} Stevens, I. R., Acreman, D. M., Ponman, T. J. 1999, MNRAS, 310, 663
 \reference{} Sun, M., \& Vikhlinin, A. 2005, ApJ, 621, 718 (SV05)
 \reference{} Tamura, T. et al. 1998, PASJ, 50, 195
 \reference{} Vollmer, B., Cayatte, V., Balkowski, C., Duschl, W. J. 2001, ApJ, 561, 708
 \reference{} Wang, Q. D., Owen, F., Ledlow, M. 2004, ApJ, 611, 821
 \reference{} Woudt, P. A., Kraan-Korteweg, R. C., Fairall, A. P. 1999, A\&A, 352, 39
 \reference{} Woudt, P. A. et al. 2004, A\&A, 415, 9

\end{references}
\end{document}